\documentclass[twoside]{photon2007}
\usepackage[latin1]{inputenc}
\usepackage[dvips]{graphicx,epsfig,color}
\usepackage{wrapfig}
\usepackage{amssymb,amsmath,array}

\begin{document}
\title{
INTRODUCTION TO  PHOTON 2007   } 
\author{Maria Krawczyk
\thanks{Supported in part  by EU Marie Curie Research Training Network HEPTOOLS,
under contract MRTN-CT-2006-035505, FLAVIAnet contract No.
MRTN-CT-2006-035482, by a Maria Curie Transfer of Knowledge
Fellowship of the European Community's Sixth Framework Programme
under contract  MTKD-CT-2005-029466 (2006-2010).}
\vspace{.3cm}\\
Institute of Theoretical Physics, University of Warsaw, ul.
   Ho\.za 69, 00-681 Warsaw, Poland}

\maketitle

\begin{abstract}
The introductory remarks to the conference Photon 2007 organized at
the Sorbonne in Paris in July 2007 are presented.

\end{abstract}
\section{Outline}

The Photon 2007 conference consists of  International Conference on
the Structure and Interactions of the Photon (organized since 1994;
since 2003 it covered also interactions of the proton) including the
17-th Int. Workshop on Photon-Photon Collisions (with the first Int.
Colloquium on Photon-Photon Collisions in Electron-Positron Storage
Rings held here in Paris, in 1973) and Int. Workshop on High Energy
Photon Linear Collider (included in the PHOTON conference in 2005;
first workshop on photon colliders was organized at LBL in 1994).

Let us reflect on  a possible content of the Photon 2009. Perhaps
it will consist of:\\
$\bullet$ Int. Conference on the Structure and Interactions of the
Photon {\bf {and the Proton}}  with results from HERA, Tevatron,
LHC\\
$\bullet$  18-th Int. Workshop on Photon-Photon Collisions with
final results from LEP and possibly first results from LHC, in
particular on the high-energy photon collisions at LHC \\
$\bullet$  Int. Workshop on High Energy Photon Linear Collider
(PLC 2009) covering both the  ILC and CLIC options\\
$\bullet$ Photons in astrophysics.

Below I will review some topics presented at the Photon2005
conference, with focus on 40th anniversary of the Vector Dominance
Model, then open problems  and new ideas related to the photon as
well as the potential of the Photon Linear Collider will be
described.
\section{The Photon: its First Hundred Years and the Future}
The previous Photon 2005 conference was organized in the Year of
Physics and therefore had a special structure: The Centenary and the
PHOTON 2005 organized in Warsaw, the PLC 2005 - in Kazimierz
(Poland). Below some of review talks,
available in the proceedings~\cite{proc2005} or the web page~\cite{web}, are listed:\\
R.  Struewer -
{\it Einstein's Revolutionary Light-Quantum Hypothesis}\\
H. Kragh- {\it Let There Be Light: Cosmic Photons Prior to the
Microwave
Background  Cosmology}
 \\ A. Lawrence- {\it Multiwavelength Universe}\\
S. Haroche- {\it The modern version of the Einstein-Bohr Photon Box:
exploring the quan-tum with  atoms and photons in a cavity}
\\J. Schneider- {\it Research with Free Electron Lasers for Soft and Hard X-rays} \\
D. Schlatter- {\it CERN and the physics of the photon and its weak}
partners \\A.Wagner- {\it The two faces of the Photon}\\D. Gross-
{\it Einstein and the Quest for Unification}\\
L.Okun- {\it Photon: History, Mass, Charge}\\
N. Straumann- {\it Gauge principle and QED}\\
E. de Rafael- {\it QED precision tests},\\ and many more  reviews on
historical aspects of hadronic interaction of photons by P. Zerwas,
D. Schildknecht, A. Buras, S. Brodsky, V. Telnov, V. Fadin, I.
Ginzburg, and others.
\subsection{40 years of VDM} Vector Meson Dominance is a  40-years old,
from the Photon 2007 perspective, idea which is  still important and
useful. As described at PHOTON 2005 by D. Schildknecht
\cite{Kuroda:2005xs}, S. Sakurai advocated in 1960 to apply the
notion of conserved currents, gauge principle and universality of
couplings in describtion of  strong interaction. He predicted the
existence of vector mesons coupled to the hadronic isospin and
hypercharge currents ($\rho,\omega,\phi$), which have been then
discovered in years 1961-3. The noninvariance of the mass term of
vector mesons was ignored temporary.

These vector mesons were "found" even earlier in the description of
the formfactors of nucleons. The current-field identity (CFI) was
proposed with a  electromagnetic field  identified with a linear
combination of isovector and isoscalar vector meson fields:
$J_\mu^{em}=J_\mu^3+\frac{1}{2}J_\mu^Y$, with eg.
$J_\mu^3=-\frac{m_\rho^2}{2\gamma_\rho} \rho_\mu$. An amplitude
describing an interaction of a photon $\gamma^*$ with virtuality
$q^2$ with hadrons A, B is:
$$[\gamma^*A\to B]=-e\frac{m_\rho^2}{2\gamma_\rho}
\frac{1}{q^2-m_\rho^2}[\rho^0 A\to B]+(\omega,\phi)$$ via an
interaction of the on-shell $\rho,\omega,\phi$ states.

For elastic scattering $\gamma^*A\to A, \gamma^*B\to B$ one finds
university of the vector-meson coupling ($f_{\rho AA}=f_{\rho
BB}=...=f_\rho$, where $f_\rho=2 \gamma_\rho$), arising from the
universality of the electromagnetic coupling $e$. The coupling
constant $f_\rho$ has been measured in $e^+e^-$ annihilation, in eg.
the $\rho \to 2 \pi$ channel, in  Novosibirsk and Orsay operating
since 1966/7.

According to CFI the electromagnetic  field is due to vecor mesons,
therefore, eg.
$$\partial^\mu F^3_{\mu\nu}=\frac{em_\rho^2}{2\gamma_\rho}\rho_\nu,$$
But what is a Lagrangian? The correct Lagrangian, consistent with
gauge invariance, were proposed by Kroll,Lee,Zumino in 1967:
$$L=-\frac{e}{2f_\rho}\rho_{\mu\nu}F^{\mu\nu}+\frac{e}{f_\rho}A_\mu
J^{(\rho)\mu}$$ or
$$L'=\frac{e'}{2f_\rho}\rho'_{\mu}A'^{\mu}-\frac{1}{2}(\frac{e'}{f_\rho})^2m_\rho^2
A'^2_\mu,$$ with $e^2=e^{'2}/(1+e^{'2}/f^2_\rho)$, and similar
linear relations between unprimed and primed fields. In such way
mass of the photon is  equal 0 and  the photon propagator has at the
lowest order  two relevant contributions arising from two terms in
the $L'$.

CFI predicted "hadronlike behavior of the photon", in particular a
relation for amplitudes $A_{\gamma p \to
\rho^0p}=\frac{e}{2\gamma_\rho}A_{\rho^0 p \to \rho^0 p}$. A
diffractive peak, typical for pure hadronic processes, was
established for the vector meson photoproduction  in 60-ties (XXc.)
at DESY and SLAC.

It was in 1967 when Stodolski applying this vector meson dominance
ideas to derive a sum rule combining the forward Compton-scattering
cross-section for $\gamma p \to V p (V=\rho,\omega,\phi)$ and the
total $\gamma p$ cross-section at high energy. However, to agree
with $\sigma_{\gamma p}$ data  about 20 \% larger Compton
contribution was required. It led to the notion of the Generalized
Vector Dominance (GVD) model by Schildknecht, Sakurai (1972) with
additional contribution of continuum massive vector states, which
start to dominate for large virtuality of the photon.

In the same  1967 year the question was posted on the dependence on
mass number $A$ of the photon-nucleus interaction and then the
answer was given (i.e. hadronlike behavior of the cross section
$\approx A^{2/3}$) by Stodolsky. The further development of these
ideas, together with a shadowing phenomena, led to a deep-inelastic
scattering physics, with a treatment of vector mesons as
quark-antiquark   states.

Finally, it is worth mentioning  here a low energy effective
lagrangian approach introduced in early 80-ties (XXc.)
\cite{Bando:1984ej}, called a Hidden Local Symmetry Lagrangian,
where  the vector mesons are treated as dynamical gauge bosons of a
spontaneously broken hidden local symmetry.  This model, with
 isospin breaking effects mostly from  the $\rho,\omega,\phi$
mixing, was used very recently \cite{Benayoun:2007cu} to describe
properly both the pion formfactor data in the $e^+e^-$ annihilation
and in the $\tau$ decay. This is important for a reliable estimation
of the vacuum polarization contribution for the $g-2$ for a muon,
see below.

\subsection{Curve and photons}
\begin{figure}[h]
\centerline{\includegraphics[width=1.0\columnwidth]{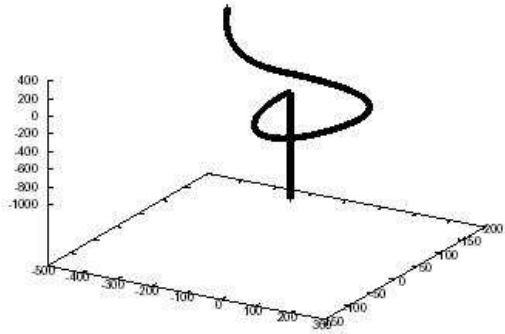}}
\caption{A curve which number is n=38.8.}\label{Fig:krzywa}
\end{figure}

Less standard idea was presented at PHOTON 2005 by L. Stodolsky
\cite{Stodolsky:2005hj}.
 If the initial and final velocity of the radiating particle are the
same  there is no IR catastrophe and
 the  number of photon  radiated by a charge  following a given
curve is finite. This number can be used to characterize the curve,
 as  it is shown in Figure~
\ref{Fig:krzywa}, for a curve which number is n=38.8.
\section{Open problems}
Here I will mention  two new  results, related to the photon-hadron
interaction, which are relevant for a search for a new  physics.
They may help to close the for-long open problems. Then some old-new
ideas related to the photons will be presented as well as the photon
connection to the Higgs and Dark Matter physics.
\subsection{Photons and hadrons}
\paragraph{$\bullet$Hadronic contribution to (g-2)$_\mu$.}
The hadronic contributions to the SM anomalous magnetic moment (g-2)
for the muon \cite{Miller:2007kk},  the light-by-light (lbl)
\cite{Vainshtein:2007zz} and the vacuum polarisation,  give the
highest uncertainties in the SM result. Moreover, for the vacuum
polarisation, where low-energy contribution has to be obtained using
some experimental data, there is an discrepancy between the
estimation based on  the $e^+e^-$  and on the $\tau$-decay data, as
presented in Figure~{\ref{Fig:g2} \cite{Jegerlehner:2007xe}. As more
straightforward are the $e^+e^-$ data, with a dominant contribution
due to the $\pi \pi$ channel,  recently only these data are being
used to derive the SM prediction. However very recently it was shown
\cite{Benayoun:2007cu} that the model based on the Hidden Local
Symmetry Lagrangian \cite{Bando:1984ej}, mentioned above, provides
for a first time the consistent  description of both the $e^+e^-$
and the $\tau$-dipion spectra. It confirms therefore ~3.2-3.3
$\sigma$ disagreement between the SM prediction and experimental
data on anomalous magnetic moment for muon \cite{Bennett:2006fi}.

\begin{figure}[h]
\centerline{\includegraphics[width=1.0\columnwidth]{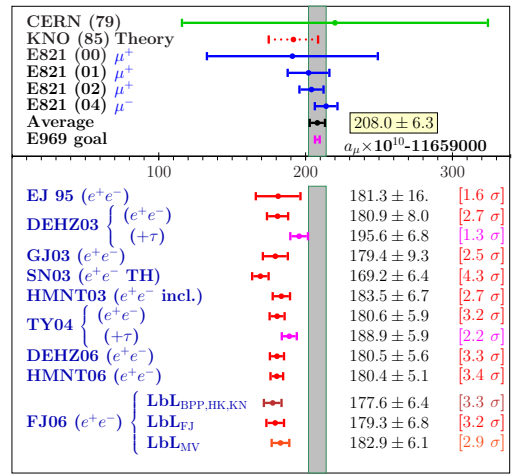}}
\caption{SM-prediction and data for $a_\mu$
\cite{Jegerlehner:2007xe}.}\label{Fig:g2}
\end{figure}

\paragraph{$\bullet$ Production of $b\bar b$ in $\gamma \gamma$ collision.}
The new ALEPH \cite{:2007ak} data on the production of the $b \bar
b$ pairs in $\gamma \gamma$ collision show an agreement with QCD, in
contradiction with three other LEP experiments. Here the $b$ quarks
were identified using lifetime information and the cross section for
$\sqrt s$ = 130 - 209 GeV was found to be $\sigma(e^+ e^- \to e^+
e^- b \bar{b} X) = (5.4 \pm 0.8 (stat) \pm 0.8 (syst))$ pb  (NLO QCD
prediction is 2.1 - 4.5 pb). The only other published LEP results on
$b$-quark production in $\gamma \gamma$ collisions is by the L3
Collaboration, obtained from a fit to the transverse momentum of
leptons with respect to jets. The cross section was measured to be
about three times the prediction of NLO QCD and similar results have
been reported at conferences by OPAL  and DELPHI.

\subsection{Light from the hidden sector}
There is a very old but still actively tested idea, which appears in
the extensions of SM, on a possible existence of a second species of
photon which is uncoupled to known forms of matter -  the {\it
hidden} (sector) photons. In 1982 Okun proposed a para-photon model
(other name {\it exphoton})\cite{Okun:1982xi}, with mixing of
massive hidden photon with the ordinary photon. It appears in the
QED if an extra U(1) symmetry is introduced, with the {\it
para-photon} as a gauge boson  and with a corresponding {\it
para-charge}. Since in such models a very light charged scalar may
appear and no such particle is observed, it is expected that the
electric charge of this particle must be very small fraction of an
electron electric charge ({\sl {mini-charge or
milli-charge}})~\cite{Holdom:1985ag}. Similar model with oscillation
of photons was considered in \cite{Georgi:1983sy}. For hidden sector
photons one can search for in the precision optical experiments, in
particular light-shining-through-walls
experiments~\cite{Ahlers:2007rd}. Also  the Super-Kamiokande
\cite{Gninenko:2008jz} can be sensitive to the hidden sector photons
as they can be produced through oscillation from the photons emitted
by Sun. It is worth mentioning that milli-charged particles may
influence cosmic microwave background
radiation~\cite{Melchiorri:2007sq}. It  can offer the explanation of
the galactic 511 keV line as coming from MeV milli-charged dark
matter~\cite{Huh:2007zw}.

\begin{figure}[h]
\centerline{\includegraphics[width=1.0\columnwidth]{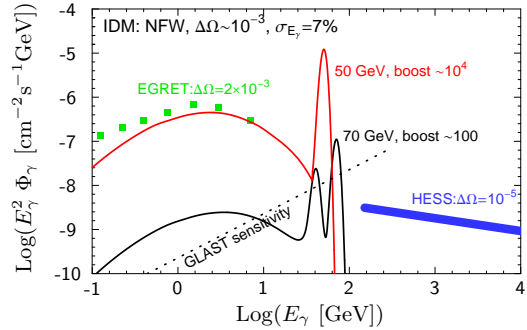}}
\caption{The energy spectrum of the photons from the annihilation
$H_D H_D$ (mass 50 and 70 GeV) to the $\gamma \gamma$; the limits
from EGRET and HESS and sensitivity from the GLAST experiments
\cite{Gustafsson:2007pc}. }\label{Fig:lines}
\end{figure}

\begin{figure}[h]
\centerline{\includegraphics[width=1.0\columnwidth]{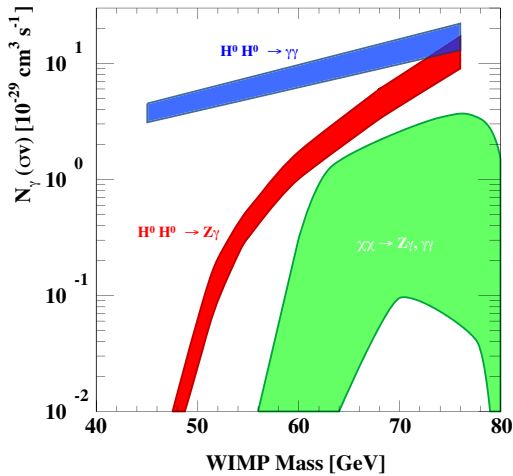}}
\caption{Comparison of the rate for annihilation of dark scalar $H_D
H_D$ and neutralino $\chi \chi$ to the $\gamma \gamma, Z \gamma$
final states as a function of the mass of dark matter candidate
\cite{Gustafsson:2007pc}.}\label{Fig:glast}
\end{figure}

\subsection{Photons, Higgs and dark matter}
\paragraph{$\bullet$Photons and Higgs sector at colliders.} Photons play an important role
in testing the Standard Model,  in particular its Higgs sector.
Higgs mechanism in this model relies on introducing one doublet of
scalar fields and spontaneous EW symmetry breaking.  As a result the
W and Z gauge bosons become massive, while the photon remains
massless. An existence of one spin-0 neutral Higgs particle $h$ is
predicted. According to the LEP data, its mass should be  above
114.4 GeV. Neutral Higgs boson can couple with photons only via a
loop with contribution from all charged particles of the theory. The
unique property of this coupling is that heavy particles, which get
masses from the Higgs mechanism, do not decouple.

In the simplest extension of the SM with two doublets of scalar
fields (2HDM, MSSM) five Higgs bosons appear,  two charged $H^\pm$
and three neutral  $h_{1,2,3}$. (In the CP-conserving neutral scalar
sector there are two CP-even Higgs bosons $h,H$ and one CP-odd $A$.
If CP is violated three neutral Higgs bosons $h_i$ mix.) CP property
or possible mixing of the neutral Higgs bosons can be tested in the
$\gamma \gamma$ collisions. Also for charged Higgs bosons  search
photons can play crucial role  due to their direct couplings.

At the LHC, the important channel for a search for a light SM-like
Higgs boson is  $gg\to Higgs \to \gamma \gamma$.  Note, however that
the intermediate mass Higgs-boson may decays into a photon and the
unparticle~\cite{Georgi}(via a one-loop process) with the branching
ratio compared with decay to two photons ~\cite{Cheung:2007sc}. At
the LHC   there is also a possibility to produce Higgs boson in the
$\gamma \gamma$ collision. However, such production process  can be
studied more precisely  at the Photon Linear Collider (PLC), a
possible option of the International Linear Collider (ILC) and of
the CLIC, with the energy range 0.5-1 TeV and 3-4 TeV, respectively.

\paragraph{$\bullet$Significant $\gamma$-lines in Inert (Dark) 2HDM.}
If  to the SM-type scalar doublet $\phi_1$ another scalar doublet
with an odd $Z_2$-parity (for $Z_2$ transformation $\phi_2 \to
-\phi_2$) is added one gets the $Z_2$-conserving Higgs sector only
if $\phi_2$ has zero vacuum expectation values (vev) and does not
couple directly to the fermions.
In such case  Higgs sector consists of  the additional with respect
to $h$ four $Z_2$-odd scalars ($H_D,A_D,H^\pm _D$)  and the lightest
neutral one, eg. $H_D$, can be a good candidate for the Dark Matter.
Such model, called the Inert (Dark) Doublet Model,  was introduced
by Deshpande and Ma in 1978~\cite{Deshpande:1977rw}, later it was
considered also by other authors, for example in
~\cite{Barbieri:2005kf,LopezHonorez:2006gr}, both from point of view
of particle physics and astrophysics.  It was found that for the
$H_D$ mass between 40 and 80 GeV the correct cosmic abundance is
obtained (WMAP).

In \cite{Gustafsson:2007pc} possibility of a significant gamma lines
from the annihilation $H_D H_D \to \gamma \gamma, \,\, Z \gamma$ was
investigated. This loop-induced mono-chromatic $\gamma$ production
would be exceptionally strong, Figure~\ref{Fig:lines}. It was shown
that these events would be ideal to search for in the upcoming GLAST
experiment. Comparison of the rate for these events and for events
with neutralino dark matter candidates is shown in
Figure~{\ref{Fig:glast}}.

\section{Photon Linear Collider - PLC}
\subsection{PLC - an option at ILC}
The project called the International Linear Collider (ILC),
corresponds to the planned $e^+e^-$ collider with energy  0.5 - 1
TeV. It is described in detail in the  ILC Reference Design
Report~\cite{Brau:2007zza} released in August 2007. The $\gamma
\gamma$ and $e\gamma$  options, called in short the Photon Linear
Collider (PLC),  can be realized at the ILC by using backward
Compton scattering of the electron beam on a laser light
\cite{Badelek:2001xb}. The PLC is not a part of the baseline ILC
design and  according to the David J. Miller, who has chaired the
Gamma-Gamma Planners (GGP) meeting during PLC2005, following issues
need to be tackled in order to make PLC a reality: optical cavity,
beam dump, luminosity maintenance, crossing angle, backgrounds. Also
a list of "golden" processes \cite{Boos:2000ki} whose study will
justify the PLC option needs to be reviewed.

In the PLC option both energy and polarization of the photon beams
vary, since they are produced in the scattering process. One can
choose polarization of the electron beam (for PLC only electron
beams are needed) and laser light in such way to get monochromatic
highly polarized photon beam. In particular one can have high
luminosity $\gamma \gamma$ option corresponding to the high-energy
peak ranging from 0.6 to 0.8 of the energy of the parent $e^+e^-$
collision. In such option a resonance production of C=+ states (eg.
Higgs boson) allows to make very precise determination of its
properties. Both $\gamma \gamma$ and $e \gamma$ have  higher mass
reach than the corresponding $e^+e^-$ collider, since here a single
production is possible. High polarization of the beams (both
circular and linear) allows to treat PLC$_{\gamma \gamma}$ as a CP
filter, since two photons can form a $J_z=0$ state, both with  even
and odd CP parities.

The physics potential of the PLC is very reach. It is an ideal
observatory of the scalar sector of the SM and beyond, leading to
important and in many cases complementary to the $e^+e^-$ILC case
tests of the EW symmetry breaking mechanism.
It is also  best place to study  hadronic interaction of the photon,
both in $\gamma \gamma$ and $e \gamma$ options, for a really real
(ie. not Weizs\"acker-Williams) high-energy photons.

\begin{figure}[h]
\centerline{\includegraphics[width=1.0\columnwidth]
{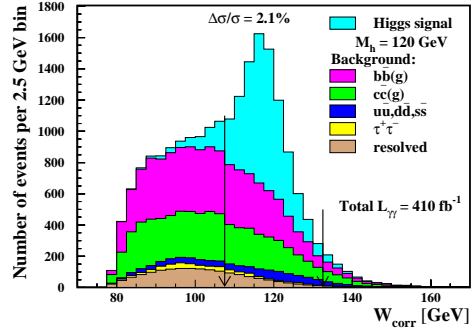}} \caption{ Distributions of the corrected
invariant mass, $W_{corr}$, for selected $b \bar{b}$ events;
contributions of the signal for $M_h = $ 120~GeV and of the
background processes \cite{nzk}.}\label{Fig:SM}
\end{figure}

\begin{figure}[h]
\centerline{\includegraphics[width=1.0\columnwidth]
{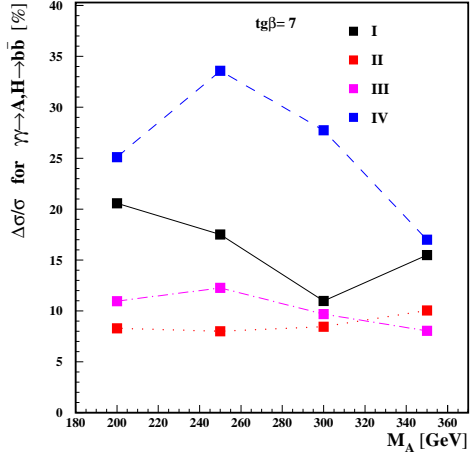}} \caption{Precisions of $\sigma (\gamma
\gamma \rightarrow A,H \rightarrow b \bar{b})$ measurement expected
after one year of the PLC$_{\gamma \gamma}$ (TESLA) running, for
$M_A = $ 200--350~GeV, $\tan \beta =$ 7 and  four MSSM parameter
sets \cite{Niezurawski:2006ia}.}\label{Fig:precmssm}
\end{figure}

Independently whether at the LHC the  SM-like  Higgs scenario will
be found or  completely new phenomena will be discovered, PLC may be
useful to clarify a picture.

\subsection{9 good reasons to build a PLC}

\paragraph{1.Precise measurement of the  $h\gamma \gamma$ coupling.}
 The  s-channel resonance production of  C=+ particle  allows to
perform precise measurement of its properties. The precision of the
cross-section measurement for the SM Higgs decaying into $b\bar b$
final state is between 2 to 3 \% for mass 120-155 GeV (Figure
~\ref{Fig:SM}) \cite{nzk,Monig:2007py}. By combining this production
rate with the  ~1~\% accuracy measurement of the $Br(h\rightarrow
bb)$ at the $e^+e^-$ ILC,  the width $\Gamma(h\rightarrow \gamma
\gamma)$ can be determined with  accuracy 2 \%, for mass of 120 GeV.
This allows to discriminate even between the SM-like Higgs models.
For mass range 200-350 GeV accuracy of the $\gamma \gamma \to WW$
cross-section measurement is still high: 3 - 8 \%. Due to the
interference with non-resonant background in WW/ZZ
~\cite{Niezurawski:2002jx} and $t \bar t$
channels~\cite{Asakawa:2003dh} it is possible to measure not only
absolute value of the $h \gamma \gamma$ amplitude but also its
phase.

\paragraph{2.Testing the Higgs-self coupling.} Production in the  $\gamma \gamma$
collision pairs of neutral Higgs bosons allows to test the trilinear
Higgs couplings, necessary for a reconstruction of the Higgs
potential. They can be measured for the Higgs-mass range 120-150 GeV
at lower energies than at the $e^+e^-$ ILC and with higher precision
than at the LHC and ILC \cite{Belusevic:2004pz}.

\paragraph{3.Covering the  LHC wedge.}
PLC can play important role in covering the so called LHC wedge,
which appears for the MSSM  for the intermediate $\tan \beta$. For
these parameters  LHC  may not be able to discover other Higgs
particles beside the lightest SM-like Higgs boson $h$. Also at the
ILC (with CMS energy  500 GeV) it may not be possible, while
PLC$_{\gamma \gamma}$ option of such collider  an observation of
heavy (degenerate) A and H bosons, with masses above 200 GeV, would
be feasible (Figure~\ref{Fig:precmssm})
\cite{Asner:2001ia,Niezurawski:2006ia,Spira:2006aa}.
\paragraph{4.Testing CP properties.}
Testing the CP nature of the Higgs bosons can be performed at the
PLC by using the initial polarization asymmetries and/or  from the
observation of decay products \cite{Grzadkowski:1992sa}. For the ZZ
and WW decays channels the angular distribution of the secondary WW
and ZZ decays products can be used \cite{Niezurawski:2004ga}. In the
$\gamma \gamma \rightarrow Higgs \rightarrow \tau\bar \tau /t \bar
t$ one can perform a model independent study of the CP-violation,
exploiting fermion polarization
\cite{Asakawa:2000jy,Asakawa:2003dh,Godbole:2006eb}. Higgs formation
in $\gamma \gamma$ collisions proves particularly interesting for
observing effects of the H/A mixing \cite{Choi:2004kq} and for CPX
scenario of MSSM \cite{Ellis:2004hw,Accomando:2006ga}, in particular
to look for the light CP-violating Higgs, which may escape discovery
both at the LEP and LHC.  It is feasible to perform H and A
discrimination for the LHC wedge using linear polarization of the
photons \cite{Zarnecki:2007he}.

\paragraph{5.Production of heavy sfermions in $e\gamma$.}
The $e \gamma$ option of the PLC  allows to study associated
production of heavy sfermions and light charginos/neutralinos in a
case when the $e^+e^-$ ILC energy will be not high enough for the
heavy sfermion pair production \cite{Datta:2002mz}.

\paragraph{6.Complementarity to ILC and LHC.}
Due to different coupling combinations for the Higgs bosons or SUSY
particles production precision measurements at $pp$, $e^+e^-$ and
$\gamma\gamma$ collisions  give   complementary  information
allowing to differentiate between various  models. In
Figure~\ref{Fig:synergy-cp},   a comparison of
   determination of the relative couplings to gauge
 bosons and top quark as well as the CP mixing parameter (angle $\Phi_{HA}$)
in the CP violating 2HDM  at LHC, ILC and PLC$_{\gamma \gamma}$
\cite{Niezurawski:2006hy} is presented.

\begin{figure}
  \hskip -0.5cm
  \includegraphics[width=8cm,angle=0]{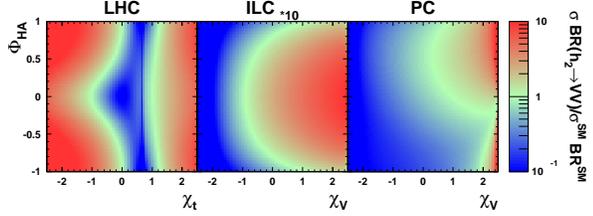}
  \caption{A
   determination of the relative couplings to $V=W/Z$ and $t$-quark as well
   as a CP mixing angle $\Phi_{HA}$
in the CP violating 2HDM  at LHC, ILC and PLC$_{\gamma \gamma}$
\cite{Niezurawski:2006hy}.}\label{Fig:synergy-cp}
\end{figure}

\paragraph{7.Photon structure and QCD tests.}
 By combining results of the dedicated $\gamma \gamma$ and $e\gamma$ measurements
quark and gluon distributions in the photon can be precisely
measured over the wide kinematic range. Also the measurement of the
spin dependent structure functions of the photon is possible. The
total cross section for $\gamma \gamma \rightarrow$ hadrons is  of
fundamental importance.

\paragraph{8.Anomalous W and $t$ couplings.}
The  cross sections for the $\gamma \gamma \rightarrow W^+W^-$ and
$e^- \gamma \rightarrow \nu W^-$ processes are very high at PLC
allowing to study the anomalous $WW\gamma$ coupling with accuracy
similar to that in $e^+e^-$ collider. At the PLC$_{\gamma \gamma}$
there is a large sensitivity for anomalous $tt\gamma$ couplings, due
to the 4th power dependence of it of the $t\bar t$ production rate.
Note, that at $e^+e^-$ ILC the couplings $\gamma t \bar t$ and $Ztt$
enter together.  The single t production at the PLC$_{e\gamma}$  is
the best option to measure the $Wtb$ coupling.

\paragraph{9.New physics in $\gamma \gamma \rightarrow \gamma \gamma /
\gamma Z / ZZ$.}
 Neutral gauge boson production,
$\gamma \gamma \rightarrow \gamma \gamma / \gamma Z / ZZ$, which
appear only at the  one-loop level in the Standard Model, allows to
derive strong constraints on new physics contributions
\cite{Gounaris:1995mc,Gounaris:2000rs}, especially if they rely on a
direct coupling to photons like for unparticles. In the $\gamma
\gamma \to \gamma\gamma$
~\cite{Cakir:2007xb,Chang:2008mk,Kikuchi:2008pr} the peculiar phases
associated with the unparticle propagator in the
 s-channel give rise to unusual patterns of interference with t- and
 u-channels.
Large deviation from the SM results (which are symmetric in s, t and
u) can be obtained at PLC$_{\gamma\gamma}(500)$ even for the cutoff
scale 5 TeV.

\section{Acknowledgments}
I wish to thank organizers for interesting and important meeting and
support. Work on this contribution was done to large extent at
Theory Group at CERN and I would like to thank CERN Theory Division
for a very warm hospitality and support. I am grateful to Peter
Zerwas, Arkady~Vainshtein and Bolek Pietrzyk for useful discussions.


\begin{footnotesize}


\end{footnotesize}

\begin{thebibliography}{99}




\bibitem{url} Slides: \\
\verb$http://indico.cern.ch/materialDisplay.py?$ \\
\verb$contribId=2&sessionId=7$ \\
\verb$&materialId=slides&confId=3841$
\bibitem{proc2005} Acta Physica Polonica 37 (2006) vol3/4 and e-Proceedings
\bibitem{web} http://photon2005.fuw.edu.pl/photon2005/
\bibitem{Kuroda:2005xs}
  M.~Kuroda and D.~Schildknecht,
  Acta Phys.\ Polon.\  B {\bf 37}, 835 (2006)
  [arXiv:hep-ph/0511091].
\bibitem{Bando:1984ej}
  M.~Bando, T.~Kugo, S.~Uehara, K.~Yamawaki and T.~Yanagida,
  Phys.\ Rev.\ Lett.\  {\bf 54}, 1215 (1985).
\bibitem{Benayoun:2007cu}
  M.~Benayoun, P.~David, L.~DelBuono, O.~Leitner and H.~B.~O'Connell,
  arXiv:0711.4482 [hep-ph].
\bibitem{Stodolsky:2005hj}
  L.~Stodolsky,
  Acta Phys.\ Polon.\  B {\bf 37}, 977 (2006)
  [arXiv:math-ph/0511085].
\bibitem{Miller:2007kk}
  J.~P.~Miller, E.~de Rafael and B.~L.~Roberts,
  Rept.\ Prog.\ Phys.\  {\bf 70}, 795 (2007)
  [arXiv:hep-ph/0703049].
\bibitem{Vainshtein:2007zz}
  A.~Vainshtein,
  Nucl.\ Phys.\ Proc.\ Suppl.\  {\bf 169}, 232 (2007).
\bibitem{Jegerlehner:2007xe}
  F.~Jegerlehner,
  Acta Phys.\ Polon.\  B {\bf 38}, 3021 (2007)
  [arXiv:hep-ph/0703125].
  \bibitem{Bennett:2006fi}
  G.~W.~Bennett {\it et al.}  [Muon G-2 Collaboration],
  Phys.\ Rev.\  D {\bf 73}, 072003 (2006)
  [arXiv:hep-ex/0602035].
\bibitem{:2007ak}
  S.~Schael {\it et al.}  [ALEPH Collaboration],
  JHEP {\bf 0709}, 102 (2007)
  [arXiv:0706.3150 [hep-ex]].
\bibitem{Okun:1982xi}
  L.~B.~Okun,
  Sov.\ Phys.\ JETP {\bf 56}, 502 (1982)
  [Zh.\ Eksp.\ Teor.\ Fiz.\  {\bf 83}, 892 (1982)].
 \bibitem{Georgi:1983sy}
  H.~Georgi, P.~H.~Ginsparg and S.~L.~Glashow,
  Nature {\bf 306}, 765 (1983).
\bibitem{Holdom:1985ag}
  B.~Holdom,
  Phys.\ Lett.\  B {\bf 166}, 196 (1986).
\bibitem{Ahlers:2007rd}
  M.~Ahlers, H.~Gies, J.~Jaeckel, J.~Redondo and A.~Ringwald,
  Phys.\ Rev.\  D {\bf 76}, 115005 (2007)
  [arXiv:0706.2836 [hep-ph]].
\bibitem{Gninenko:2008jz}
  S.~N.~Gninenko,
  arXiv:0802.1315 [hep-ph].
\bibitem{Melchiorri:2007sq}
  A.~Melchiorri, A.~Polosa and A.~Strumia,
  Phys.\ Lett.\  B {\bf 650}, 416 (2007)
  [arXiv:hep-ph/0703144].
\bibitem{Huh:2007zw}
  J.~H.~Huh, J.~E.~Kim, J.~C.~Park and S.~C.~Park,
  arXiv:0711.3528 [astro-ph].
\bibitem{Georgi:2007ek}
  H.~Georgi,
  Phys.\ Rev.\ Lett.\  {\bf 98}, 221601 (2007)
  [arXiv:hep-ph/0703260],
  Phys.\ Lett.\  B {\bf 650}, 275 (2007)
  [arXiv:0704.2457 [hep-ph]].
\bibitem{Cheung:2007sc}
  K.~Cheung, C.~S.~Li and T.~C.~Yuan,
  arXiv:0711.3361 [hep-ph].
\bibitem{Deshpande:1977rw}
  N.~G.~Deshpande and E.~Ma,
  Phys.\ Rev.\  D {\bf 18}, 2574 (1978).
\bibitem{Barbieri:2005kf}
  R.~Barbieri and L.~J.~Hall,
  arXiv:hep-ph/0510243.
\bibitem{LopezHonorez:2006gr}
  L.~Lopez Honorez, E.~Nezri, J.~F.~Oliver and M.~H.~G.~Tytgat,
  JCAP {\bf 0702}, 028 (2007)
  [arXiv:hep-ph/0612275].
\bibitem{Gustafsson:2007pc}
  M.~Gustafsson, E.~Lundstrom, L.~Bergstrom and J.~Edsjo,
  Phys.\ Rev.\ Lett.\  {\bf 99}, 041301 (2007)
  [arXiv:astro-ph/0703512].

\bibitem{Brau:2007zza}
  J.~Brau {\it et al.},
  ``International Linear Collider reference design report. 1: Executive
  summary. 2: Physics at the ILC. 3: Accelerator. 4: Detectors,''
\bibitem{Badelek:2001xb}
  B.~Badelek {\it et al.}  [ECFA/DESY Photon Collider Working Group],
  Int.\ J.\ Mod.\ Phys.\  A {\bf 19}, 5097 (2004)
  [arXiv:hep-ex/0108012].
\bibitem{Boos:2000ki}
  E.~Boos {\it et al.},
  Nucl.\ Instrum.\ Meth.\  A {\bf 472}, 100 (2001)
  [arXiv:hep-ph/0103090].

\bibitem{nzk}
  P.~Niezurawski, A.~F.~Zarnecki and M.~Krawczyk,
  Acta Phys.\ Polon.\  B {\bf 34}, 177 (2003)
  [arXiv:hep-ph/0208234],
  arXiv:hep-ph/0307183,
\bibitem{Monig:2007py}
  K.~Monig and A.~Rosca,
  arXiv:0705.1259 [hep-ph].

\bibitem{Niezurawski:2002jx}
  P.~Niezurawski, A.~F.~Zarnecki and M.~Krawczyk,
  JHEP {\bf 0211}, 034 (2002)
  [arXiv:hep-ph/0207294].
\bibitem{Asakawa:2003dh}
  E.~Asakawa and K.~Hagiwara,
  Eur.\ Phys.\ J.\  C {\bf 31}, 351 (2003)
  [arXiv:hep-ph/0305323].
\bibitem{Belusevic:2004pz}
  R.~Belusevic and G.~Jikia,
  Phys.\ Rev.\  D {\bf 70}, 073017 (2004)
  [arXiv:hep-ph/0403303].
\bibitem{Asner:2001ia}
  D.~M.~Asner, J.~B.~Gronberg and J.~F.~Gunion,
  Phys.\ Rev.\  D {\bf 67}, 035009 (2003)
  [arXiv:hep-ph/0110320].
\bibitem{Niezurawski:2006ia}
  P.~Niezurawski, A.~F.~Zarnecki and M.~Krawczyk,
  Acta Phys.\ Polon.\  B {\bf 37}, 1187 (2006).
\bibitem{Spira:2006aa}
  M.~Spira, P.~Niezurawski, M.~Krawczyk and A.~F.~Zarnecki,
  Pramana {\bf 69}, 931 (2007)
  [arXiv:hep-ph/0612369].
\bibitem{Grzadkowski:1992sa}
  B.~Grzadkowski and J.~F.~Gunion,
  Phys.\ Lett.\  B {\bf 294}, 361 (1992)
  [arXiv:hep-ph/9206262].
\bibitem{Niezurawski:2004ga}
  P.~Niezurawski, A.~F.~Zarnecki and M.~Krawczyk,
  Acta Phys.\ Polon.\  B {\bf 36}, 833 (2005)
  [arXiv:hep-ph/0410291].
\bibitem{Choi:2004kq}
  S.~Y.~Choi, J.~Kalinowski, Y.~Liao and P.~M.~Zerwas,
  Eur.\ Phys.\ J.\  C {\bf 40}, 555 (2005)
  [arXiv:hep-ph/0407347].
\bibitem{Ellis:2004hw}
  J.~R.~Ellis, J.~S.~Lee and A.~Pilaftsis,
  Nucl.\ Phys.\  B {\bf 718}, 247 (2005)
  [arXiv:hep-ph/0411379].
\bibitem{Accomando:2006ga}
  E.~Accomando {\it et al.},
  arXiv:hep-ph/0608079.
\bibitem{Zarnecki:2007he}
  A.~F.~Zarnecki, P.~Niezurawski and M.~Krawczyk,
  arXiv:0710.3843 [hep-ph].
\bibitem{Datta:2002mz}
  A.~Datta, A.~Djouadi and M.~Muhlleitner,
  Eur.\ Phys.\ J.\  C {\bf 25}, 539 (2002)
  [arXiv:hep-ph/0204354].
\bibitem{Asakawa:2000jy}
  E.~Asakawa, S.~Y.~Choi, K.~Hagiwara and J.~S.~Lee,
  Phys.\ Rev.\  D {\bf 62}, 115005 (2000)
  [arXiv:hep-ph/0005313].
\bibitem{Godbole:2006eb}
  R.~M.~Godbole, S.~Kraml, S.~D.~Rindani and R.~K.~Singh,
  Phys.\ Rev.\  D {\bf 74}, 095006 (2006)
  [Erratum-ibid.\  D {\bf 74}, 119901 (2006)]
  [arXiv:hep-ph/0609113].
\bibitem{Niezurawski:2006hy}
  P.~Niezurawski, A.~F.~Zarnecki and M.~Krawczyk,
  Acta Phys.\ Polon.\  B {\bf 37}, 1173 (2006),
\bibitem{Gounaris:1995mc}
  G.~J.~Gounaris, J.~Layssac and F.~M.~Renard,
  Z.\ Phys.\  C {\bf 69}, 505 (1996)
  [arXiv:hep-ph/9505430].
\bibitem{Gounaris:2000rs}
  G.~J.~Gounaris,
  Nucl.\ Instrum.\ Meth.\  A {\bf 472}, 181 (2001)
  [arXiv:hep-ph/0008170].
\bibitem{Cakir:2007xb}
  O.~Cakir and K.~O.~Ozansoy,
  arXiv:0712.3814 [hep-ph].
\bibitem{Chang:2008mk}
  C.~F.~Chang, K.~Cheung and T.~C.~Yuan,
  arXiv:0801.2843 [hep-ph].
\bibitem{Kikuchi:2008pr}
  T.~Kikuchi, N.~Okada and M.~Takeuchi,
  arXiv:0801.0018 [hep-ph].
\end{thebibliography}
\end{document}